\documentclass[twocolumn,showpacs,preprintnumbers,amsmath,amssymb]{revtex4}

\usepackage{graphicx}
\usepackage{dcolumn}
\usepackage{bm}

\def\hH{\hat{H}}
\def\hX{\hat{x}}
\def\hP{\hat{p}}

\def\ve{\vec{\lambda}}

\def\hm{\hat{\nu}}
\def\hn{\hat{\nu}^{\prime}}
\newcommand{\vs}{\vec{\sigma}}

\begin{document}


\title{Immeasurability of Zero-point Energy in the Cosmological Constant problem}

\author{Daegene Song}

\affiliation{%
Korea Institute for Advanced Study, Seoul 130-722, Korea
}%

\date{\today}

\begin{abstract}
A huge discrepancy between the zero-point energy calculated from
quantum theory and the observed quantity in the Universe has been
one of the most illusive problems in physics. In order to examine
the measurability of zero-point energy, we construct reference
frames in a given measurement using observables.  Careful and
explicit construction of the reference frames surprisingly reveals
that not only is the harmonic oscillator fluctuating at the ground
level, but so is the reference frame when the measurement is
realized. The argument is then extended to examine the
measurability of vacuum energy for a quantized electromagnetic
field, and it is shown that while zero-point energy calculated
from quantum theory diverges to infinity, it is not measurable.

\end{abstract}

\maketitle Vacuum, unlike its common perception as emptiness,
plays an important role in modern physics.  With the development
of quantum mechanics, it was soon discovered that, due to the
uncertainty principle, the vacuum in quantum theory is not empty,
but exists at a non-zero level of energy.  This observation
presents a significant problem, known as the cosmological constant
problem, due to the huge discrepancy between the calculation of
vacuum energy from quantum theory and the measured quantity from
the direct observation in the Universe (see
\cite{peebles,straumann} and the references therein).

Let us provide a simple example which will give an outline of the
main argument to be presented in this paper. For a particle in
one-dimensional classical harmonic oscillation, its position is
described as a function of time as follows $x(t) = \eta_1 \cos
(\omega t+\phi )$. Let us now suppose there is a detector
positioned on another harmonic oscillator that is also oscillating
with the same phase and with the amplitude $\eta_2$ as follows,
$y(t) = \eta_2 \cos(\omega t+\phi )$. If we assume $\eta_1$ and
$\eta_2$ are non-negative and $\eta_2\leq \eta_1$, then the
detector would measure the position of the particle as follows,
$x(t)-y(t) = (\eta_1-\eta_2)\cos (\omega t+\phi )$ and would
observe the energy to be $\frac{1}{2} m\omega^2
(\eta_1-\eta_2)^2$. Note that if $\eta_1$ is non-zero, then the
energy of the particle oscillation is also non-zero. However, this
value does not fully contribute to the measured quantity but only
the difference contributes to the observed energy. Also note that
when $\eta_2=\eta_1$, the observed energy is zero.

Therefore, even when the energy of the particle is not zero, if
the detector is also oscillating as discussed above, the energy
may not be measurable by the detector.  In this paper, we will
present a similar case for zero-point energy of quantum fields. In
quantum theory, observables are referred to as a quantity that can
be measured, such as position, momentum or a spin. We will
construct reference frames from observables and this will provide
us with a tool to examine the measurability of zero-point energy.
Interestingly, it will turn out that both the constructed
reference frames and the state vectors being measured could be in
one of the same possible states. Remarkably, this shows that not
only is the harmonic oscillator fluctuating at the ground level,
but so is the reference frame when the measurement is done.  Since
a quantum field may be considered as an infinite collection of
harmonic oscillators, this argument naturally generalizes to the
case of a quantum field. It will be shown that while each harmonic
oscillator has non-zero energy at the ground level, the reference
frame for each harmonic oscillator could also be sitting only at
the same non-zero energy level, thereby, the measurement outcome
only yields zero.

Let us begin by examining measurability in the case of a single
qubit. A qubit, a basic unit of quantum information
\cite{nielsen}, is a two-level quantum system written as
$|\psi\rangle =a|0\rangle + b|1\rangle$ where $a,b$ are complex
numbers.  Since our goal is to construct a measurement reference
frame and a qubit is defined in two-dimensional complex vector
space, it would be convenient to consider the qubit as an object
that is easier to be visualized, i.e., as a unit vector in the
Bloch sphere notation. Using a Bloch sphere notation, i.e., with
$a =\exp(-i\phi/2)\cos(\theta/2)$ and $b
=\exp(i\phi/2)\sin(\theta/2)$, a qubit in a density matrix form
can be written as $|\psi\rangle\langle \psi| = \frac{1}{2}({\bf
{1}}+ \hm \cdot \vs )$ where
 $({ {\hm}}_x,{ {\hm}}_y,{{\hm}}_z)$
 $=(\sin\theta \cos\phi,\sin\theta\sin\phi,\cos\theta )$ and
$\vs = (\sigma_x,\sigma_y,\sigma_z)$ where $\sigma_x =
|0\rangle\langle 1| + |1\rangle\langle 0|$, $\sigma_y =
-i|0\rangle\langle 1| + i|1\rangle\langle 0|$, and $\sigma_z =
|0\rangle\langle 0| - |1\rangle\langle 1|$. Therefore a qubit,
$|\psi\rangle\langle \psi|$, can be represented as a unit vector
${{\hm}} = ({{\hm}}_x,{{\hm}}_y,{{\hm}}_z)$ pointing in
$(\theta,\phi)$ of a sphere with $0\leq \theta \leq \pi , 0\leq
\phi \leq 2\pi$. Having the notation of writing the qubit in a
sphere, we wish to construct a reference frame in measuring the
qubit in the same sphere. In particular, we wish to use
observables in quantum theory. Similar to the case of a qubit, an
observable can also be written as a unit vector in the Bloch
sphere, ${{\hn}}=( {{\hn}}_x,{{\hn}}_y,{{\hn}}_z )$ where
$({{\hn}}_x,{{\hn}}_y,{{\hn}}_z)$ $=(\sin\vartheta
\cos\varphi,\sin\vartheta\sin\varphi,\cos\vartheta )$, pointing
$(\vartheta,\varphi)$ direction in a sphere. Therefore if one is
to make a measurement in $(\vartheta,\varphi)$ direction,
 the observable would be ${{\hn}} \cdot \vs$.

Note that using the introduced notation, both the state vector and
observables are written as unit vectors in the Bloch sphere. If we
consider the qubit to be a spin-1/2 particle pointing in some
direction $(\theta,\phi)$, the spin could be either pointing up or
down in that particular direction.  It was noted that just as the
state vector can be either spin-up or spin-down in a given
direction, $(\theta,\phi)$ in the Bloch sphere, so is the
reference frame, i.e., the vector representing the observable can
be either up or down in a given direction.  For simplicity, let us
choose the spin-1/2 particle to be pointing in $z$-direction.  As
pointed out, the particle could be either pointing in
$+z$-direction ($\hm=(0,0,1)$) or $-z$-direction ($\hm=(0,0,-1)$).
If this particle is to be measured in $z$-direction, there are two
choices of reference frame, i.e., $+z$-direction ($\hn=(0,0,1)$)
or $-z$-direction ($\hn=(0,0,-1)$). Suppose the particle is
initially prepared in spin-up $z$-direction.  If the particle is
measured with an observable pointing up in $z$-direction, the
eigenvalue outcome would be $+1$.  This then implies that the
particle's spin is in the same direction as the measurement
reference frame, $z$, and it could be concluded that the spin is
pointing up in $z$ direction. However, if the measurement
reference frame is chosen as pointing down in $z$-direction,
$\hn=(0,0,-1)$, then the eigenvalue outcome would be $-1$.
Obtaining $-1$ implies that the particle's spin is in the opposite
direction as the measurement reference frame, i.e.,  $-z$ in this
case, and therefore, it could be concluded that the particle's
spin is pointing in $+z$.  Notice that the reference frame for
measurement is not outside of spin up or down in $z$-direction and
the measurement result yields spin up or down. The reference frame
also lies either up or down just as the particle does.  The
measurement outcome only tells us if the spin is in the same
direction ($+1$) or in the opposite direction ($-1$) as the
reference frame, instead of  up or down. This observation, where
the reference frame not being outside of spin up or down, will be
important in considering the reference frame for energy
eigenvectors for harmonic oscillators. Moreover, for identically
prepared states, the measurement outcomes are opposite depending
on the choice of reference frames.  That is, the eigenvalue
outcomes $\pm 1$ are not the absolute values but they depend on
the reference frame of a given measurement. This point will be
relevant when we consider the case of harmonic oscillator that the
only the relevant quantity of energy eigenvalues will be
measurable.

\begin{figure}
\begin{center}
{\includegraphics[scale=.39]{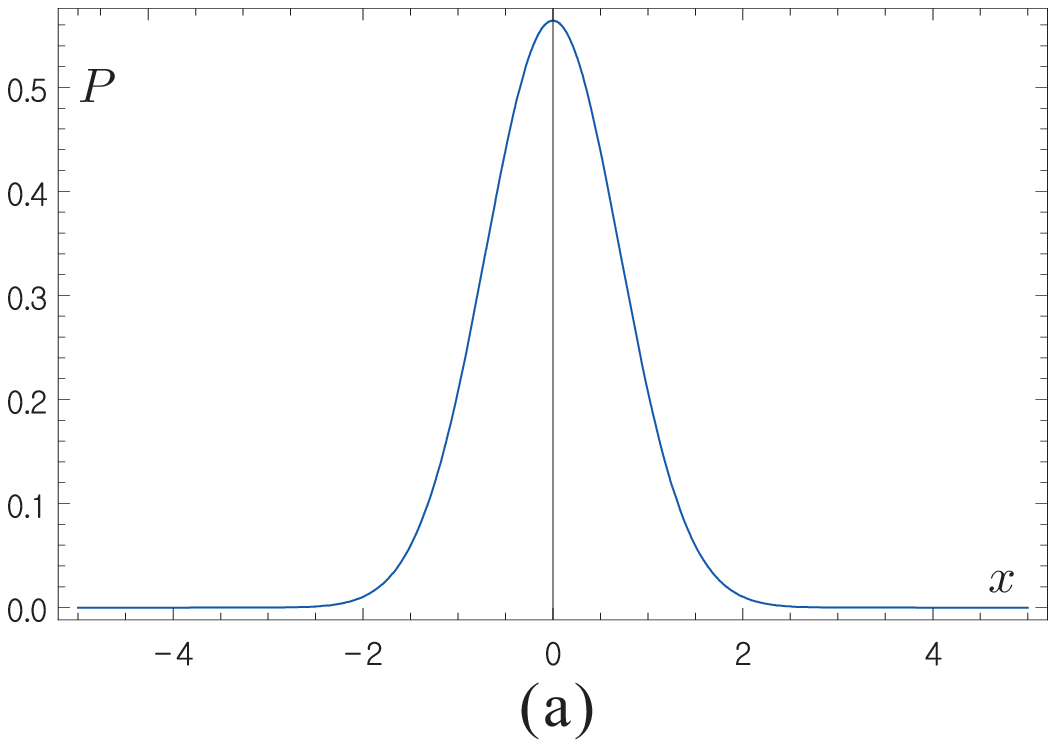}}
{\includegraphics[scale=.39]{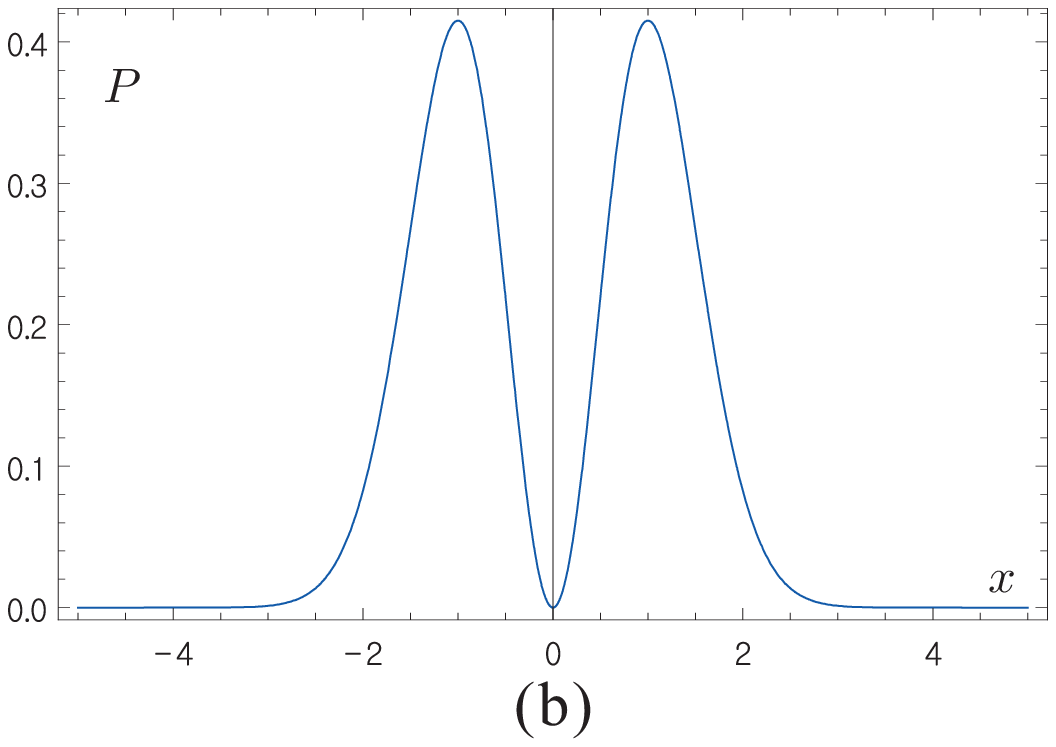} }
{\includegraphics[scale=.39]{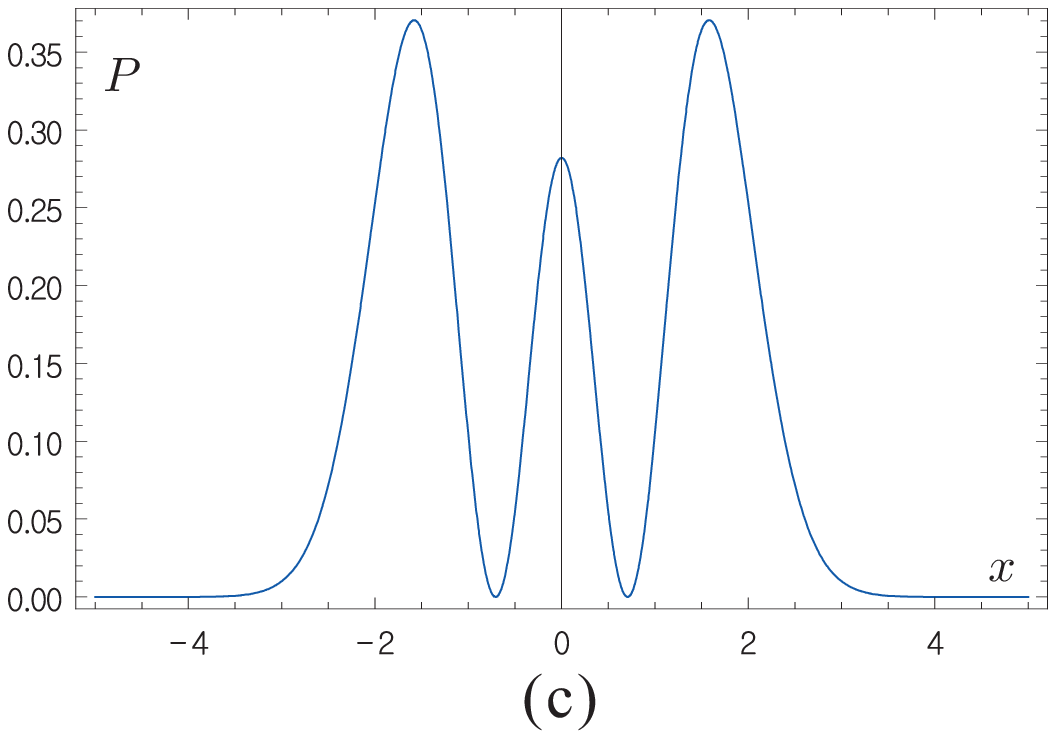}   }
{\includegraphics[scale=.39]{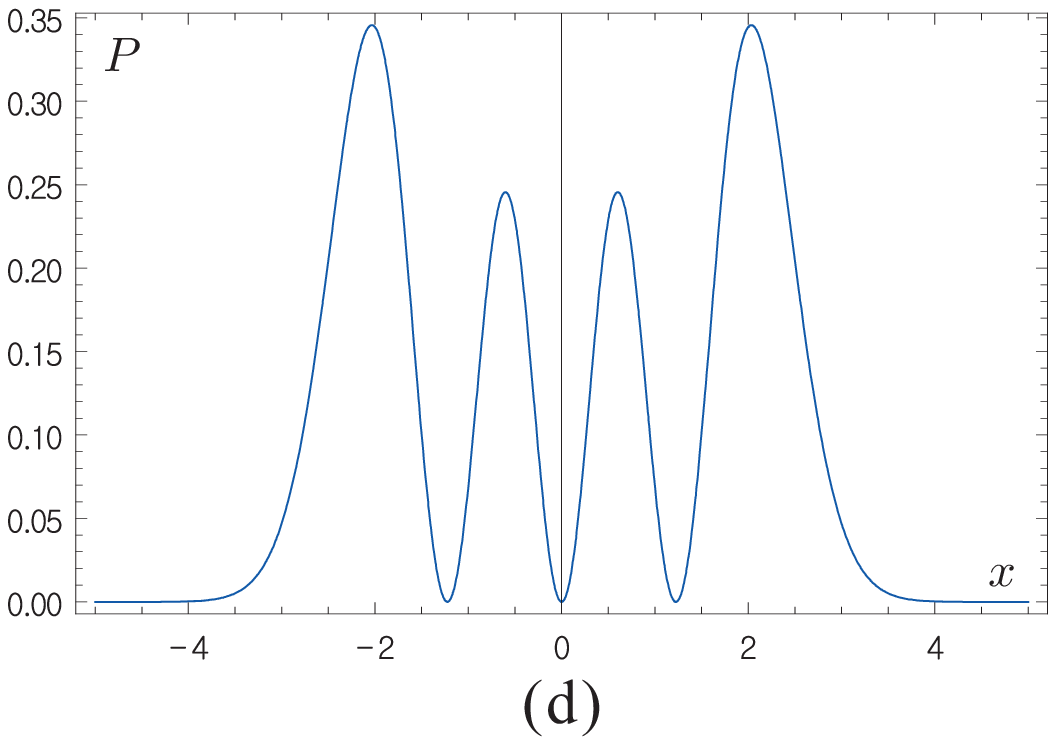} }
\end{center}
\caption{First four probability distribution of energy state of a
harmonic oscillator in the position space where we have set
$\hbar/m\omega =1$. } \label{energylevel}\end{figure}

Since a quantum field can be considered as an infinite collection
of harmonic oscillators, we wish to present our argument starting
with a single harmonic oscillator (see \cite{sakurai} for a
review). A Hamiltonian for a single one-dimensional quantum
harmonic oscillator is written in terms of position and momentum
operators as follows, $\hH = \hP^2/2m + m\omega^2 \hX^2/2$. With
the introduction of raising and lowering operators,
$a^{\dagger}=1/\sqrt{2}(\sqrt{m\omega/\hbar } \hX - i
\sqrt{1/m\hbar \omega}\hP)$ and $a =1/\sqrt{2}(\sqrt{m\omega/\hbar
} \hX - i \sqrt{1/m\hbar \omega}\hP)$, the Hamiltonian can be
re-written as $\hH = \hbar\omega(N + 1/2)$ where $N$ is a number
operator, $N=a^{\dagger} a$.  The eigenvectors of Hamiltonian can
be obtained from the following relations, $\hH
|\varepsilon_n\rangle = \varepsilon_n |\varepsilon_n\rangle$ where
$\varepsilon_n = (n+\frac{1}{2})\hbar \omega$. Let us write the
Hamiltonian, or the energy observable, in terms of its
eigenvectors as follows, $\hH=
\varepsilon_n|\varepsilon_n\rangle\langle \varepsilon_n|$. In
order to consider a measurement of the energy eigenstate, we wish
to consider the measurement in the position space. In order to
measure an energy state $|\varepsilon_n\rangle$ in the position
space, the outcome can be calculated as $|\varepsilon_n\rangle =
\int_x |x\rangle\langle x|\varepsilon_n\rangle$. Therefore, for a
given state $|\varepsilon_n\rangle$, the measurement outcome would
yield the state $|x\rangle$ with the probability, $|\langle
x|\varepsilon_n\rangle |^2$, as follows
\begin{equation}
 \frac{1}{2^n \sqrt{\pi} n! }\left( \frac{\hbar}{m\omega} \right)^{n-\frac{1}{2}} \left| \left(\frac{m\omega}{\hbar} x - \frac{d}{dx} \right)^n e^{\frac{-m\omega}{2\hbar} x^2} \right|^2
\end{equation}
The first four probability distributions are shown in Fig.
\ref{energylevel}.  When the energy state is measured in the
position space, and the probability distribution is obtained as in
(a) of Fig. \ref{energylevel}, then one can determine the energy
is in the ground state $|\varepsilon_0\rangle$. If the position
distribution comes out as seen in (b) of Fig. \ref{energylevel},
then it can be determined that the energy state corresponds to
$|\varepsilon_1\rangle$. Similarly, for (c) and (d), they
correspond to $|\varepsilon_2\rangle$ and $|\varepsilon_3\rangle$,
respectively, etc. Our goal was to construct reference frames in
measuring the energy state analogous to the qubit case. In order
to do so, let us consider the following commuting energy
observables defined as follows, $ \sum_{m=0}^{\infty}
\varepsilon_{m} |\varepsilon_{m+\ell}\rangle\langle
\varepsilon_{m+\ell}|$ where $\ell=0,1,2,...$. One can see that,
when $\ell=0$, it is identical to the Hamiltonian we studied
above. Let us consider the case when $\ell=1$. In such case, when
the energy state is $|\varepsilon_m\rangle$, the eigenvalue
obtained after a measurement would correspond to
$\varepsilon_{m-1}$.  As was discussed above, we may consider the
measurement in the position space. Therefore, when the probability
distribution is obtained as in (a) of Fig. \ref{energylevel}, then
the energy state corresponds to $|\varepsilon_1\rangle$.
Similarly, when the probability distribution is obtained as in
(b), then $|\varepsilon_2\rangle$ is the energy state and so on.
Similarly, for higher values of $\ell$ for the commuting energy
observables, measurements in the position space would yield the
determination of the corresponding energy states.

While we have provided energy states and observables, it is not
immediately clear how the commuting observables as defined above
provide reference frames in measuring the energy states. Although
measurement of a energy state $|\varepsilon_n\rangle$ in the
position space yields the probability shown in (\ref{energylevel})
(also see Fig. \ref{energylevel}), the average value of $x_n$
which corresponds to $\varepsilon_n = (n+1/2)\hbar \omega$ can be
obtained. In fact, the average of $x_n$ is zero as can be seen in
Fig. \ref{energylevel}, however, the expectation value of
${\hat{x}}^2$ yields an appropriate value for the energy
eigenvalue i.e., $\langle \varepsilon_n |{\hat{x}}^2|\varepsilon_n
\rangle = (2n+1)\hbar / 2m\omega$.  Since the indication of the
position state, along with the momentum, could represent the
energy state, we wish to introduce the notation of a position
vector $\ve_n$ to represent $(2n+1)\hbar/m\omega$. It can be
checked that the separation between two adjacent vectors $\ve_j$
and $\ve_{j+1}$ is $\Delta\lambda = 2\hbar /m\omega$ which leads
to the energy difference $\hbar \omega$ between two adjacent
levels. Therefore, the energy state can now be viewed as a
particle being at one of the equally spaced positions in a
one-dimensional coordinate starting at $\ve_0 = \hbar/m\omega$.

Let us construct reference frames in terms of the newly defined
position vectors from energy observables, $\sum_{m=0}^{\infty}
\varepsilon_{m} |\varepsilon_{m+\ell}\rangle\langle
\varepsilon_{m+\ell}|$ where $\ell=0,1,2,...$.  In case of a
qubit, we constructed reference frames from observables such that
the reference frames ended up having the same degrees of freedom
as the qubit being measured. We also noted that the eigenvalue
outcomes depended on the reference frame, i.e., when the outcome
was $+1$ ($-1$), this meant the particle's spin is in the same
(opposite) direction as the reference frame. With these two
criteria we wish to consider the observables $ \sum_{m=0}^{\infty}
\varepsilon_{m} |\varepsilon_{m+\ell}\rangle\langle
\varepsilon_{m+\ell}|$ as a detector positioned at
$(2n+1)\hbar/m\omega$, which we will denote with a position vector
$\ve_n^{\prime}$ just as in the case of eigenvectors. Let us
examine how this particular construction is consistent with the
qubit case. Note that just as a particle could be placed on one of
the equally spaced $\ve_i$'s, so is the detector. This is
analogous to the qubit case where just as particle's spin could be
up or down in a given direction, so is the reference frame in
measuring the qubit. Suppose the detector is positioned at
$\hbar/m\omega$, i.e., its reference frame corresponds to a vector
$\ve_0^{\prime}$.  When the particle is also positioned at
$\lambda_0$, i.e., at the ground level, then the outcome would
yield zero, meaning the particle's position is the same as the
detector, i.e., $\hbar/m\omega$.  Notice that while the
measurement outcome is zero, the result reveals that the particle
is positioned at $\hbar/m\omega$. When the particle is at
$\lambda_1$ and the detector is positioned at $\lambda_0$, then
the outcome would be $\Delta\lambda = 2\hbar/m\omega$.  However,
if the detector is positioned at $\lambda_1$, i.e., the reference
frame of $\ve_1^{\prime}$, measuring the same particle positioned
at $\lambda_1$ would yield zero rather than $2\hbar/m\omega$. Just
as in the qubit case, the measurement outcomes are relative, i.e.,
they depend on where the detector is positioned at.  Therefore,
representing energy states and observables through the position
vectors $\ve_i$ and $\ve_j^{\prime}$, respectively, leads us to
view the states and reference frames in a clearer picture similar
to the case of the Bloch sphere with a qubit.  That is, it
provides us to show why ground state of a harmonic oscillator is
non-zero, yet is not measurable.

We now wish to construct reference frames in measurement for
quantum fields. In particular, we wish to study the measurability
of quantized electromagnetic field in quantum optics (see
\cite{vedral} for a review). Quantizing the electromagnetic field
involves considering the Maxwell's equation in the absence of
charges. From classical Maxwell's equations in a vacuum, one could
obtain the following equation,
\begin{equation}
\nabla^2 A = \frac{1}{c^2} \frac{\partial^2 A}{\partial t^2}
\end{equation}
The solution to this has the following form, $A =  \sum_k
(u_k(t)v_k(r) + c.c.)$. With separation of variables, we could
obtain the two equations for $u_k(t)$ and $v_k(r)$, and the
solutions to these equations yield the following for the vector
potential,
\begin{equation}
A =  \sum_k \left( b_k {\bf{e}}_k e^{-i\omega t}e^{ik\cdot r} +
c.c. \right)
\end{equation}
where ${\bf{e}}_k$ is a unit polarization vector and we are
assuming the frequency $\omega$ to be the same for all $k$ modes
for simplicity. The classical Hamiltonian of the electromagnetic
field is written as
\begin{equation}
H=\frac{1}{2}\int dV (\epsilon_0 E^2 + \frac{1}{\mu_0} B^2)
\end{equation}
where $\epsilon_0$ and $\mu_0$ are electric permittivity and
magnetic permeability, respectively, that are related to the speed
of light, $c=1/\sqrt{\epsilon_0 \mu_0}$. Since $B=\nabla \times A$
and $E=-\partial A/\partial t$ and with newly defined variables,
$b_k = \sqrt{1/4 V \epsilon_0}(x_k + ip_k/\omega)$ and $b_k^* =
\sqrt{1/4 V \epsilon_0}(x_k - ip_k/\omega)$, we could obtain the
Hamiltonian in the following form,
\begin{equation}
H = \frac{1}{2} \sum_k (p_k^2 + \omega^2 x_k^2 )
\end{equation}
Quantization can be achieved by replacing $x$ and $p$ with
operators ${\hat{x}}$ and ${\hat{p}}$ which follow the usual
commutation rules. Now with the newly defined operators
\begin{eqnarray}
{\hat{a}}_k &=& \sqrt{\frac{1}{2\hbar \omega}}\left( \omega {\hat{x}}_k + i{\hat{p}}_k \right) \nonumber \\
{\hat{a}}_k^{\dagger} &=& \sqrt{\frac{1}{2\hbar \omega}}\left(
\omega {\hat{x}}_k - i{\hat{p}}_k\right)
\end{eqnarray}
the Hamiltonian can be re-written as
\begin{equation} H= \sum_k \hbar \omega ({\hat{a}}_k^{\dagger} {\hat{a}}_k + \frac{1}{2})
\end{equation}
If we define the number operator as $N_k =
{\hat{a}}_k^{\dagger}{\hat{a}}_k$, then the Hamiltonian can be
written as $H=\sum_k \hbar \omega (N_k + 1/2)$. Therefore, it can
be seen that the quantized electromagnetic field can be considered
as an infinite collection of decoupled harmonic oscillators with
frequency $\omega$ where the decoupled eigenvectors are
$|\varepsilon_{n_1}\rangle_1 \otimes |\varepsilon_{n_2}\rangle_2
\otimes \cdots $. The energy of ground state, a zero-point energy,
can be calculated as
\begin{equation}
{\mathcal{E}} =\sum \varepsilon_{0,m} = \sum_{m=1}^{\infty}
\frac{1}{2}\hbar \omega
\end{equation}
where $\varepsilon_{j,m}$ is the eigenvalue for $m$th mode
eigenvector $|\varepsilon_j\rangle_m$ and it can be seen that this
value diverges to infinity. In quantum optics \cite{vedral},
measurement of eigenvectors is done with photon number counting of
$N_k$ for each mode $k$. Therefore for a $k$th detector, when the
photon numbers are registered as $n_k$, then the energy state
corresponds to $|\varepsilon_{n_k}\rangle_k$.

In the cases of a qubit and a single harmonic oscillator, we
introduced commuting observables in order to construct reference
frame in a similar way to state vectors.  For a quantum field we
wish to do the same and introduce commuting observables as
follows, $\sum_m
\varepsilon_{m,1}|\varepsilon_{m+\ell_1}\rangle_{11}\langle\varepsilon_{m+\ell_1}|
\otimes \sum_m
\varepsilon_{m,2}|\varepsilon_{m+\ell_2}\rangle_{22}\langle\varepsilon_{m+\ell_2}|
\otimes \cdots$. When $\ell_1=0, \ell_2=0, ...$, then the
observable is the same as the Hamiltonian described above. In
order to realize this observable, i.e., where all $\ell_k$'s are
zero, when the $k$th photon number is detected as $n_k$, then the
corresponding energy state would be determined as
$|\varepsilon_{n_k}\rangle_k$. Let us now consider the case when
$\ell_1=1, \ell_2=0, \ell_3=0, ...$ This observable can be
realized by taking $|\varepsilon_{n_k}\rangle_k$ for the detection
of $n_k$ photons registered at $k$th detector except, when $k=1$,
the $n_1$ photon numbers yield $|\varepsilon_{n_1+1}\rangle_1$.
Similarly, when $\ell_1=2$ and all other $\ell_j$'s are zero, then
for $n_1$ photon number detection, the state could be determined
as $|\varepsilon_{n_1+2}\rangle_1$, while for $k\neq 1$, the
energy states could be decided the same as described above. This
analogy can be extended to higher values of $\ell_1$ and for other
$\ell_j$'s. This describes how the commuting observables can be
realized with photon number detectors.

In case of a single harmonic oscillator, we used the fact that the
average position which corresponds to the appropriate energy state
can be useful in constructing reference frames that are easy to
visualize. Therefore the energy eigenvectors for a quantum field
could be viewed as a collection of particles where each particle
is positioned at $\lambda_j = (2n_j + 1)\hbar/m\omega$ on each
coordinate. Since we have an infinite number of harmonic
oscillators, we rewrite the energy states,
$|\varepsilon_{n_1}\rangle_1 \otimes |\varepsilon_{n_2}\rangle_2
\otimes \cdots $, with the following position vectors $\ve_{n_1,1}
\otimes \ve_{n_2,2} \otimes \cdots $. We now wish to construct
reference frames from observables which have the same degrees of
freedom as position vectors.  For a single harmonic oscillator, we
argued that the commuting observables could be pictured as a
detector being positioned at $\lambda_n = (2n+1)\hbar /m\omega$.
In this way, we were able to construct reference frames of
measurement. We now wish to extend this to a multiple number of
detectors. We will now assume that a detector positioned at
$\lambda_{j}=(2j + 1)\hbar/m\omega$ on $k$th coordinate to be
denoted as $\ve_{j,k}^{\prime}$.  This then yields the commuting
observables $\sum_m
\varepsilon_{m,1}|\varepsilon_{m+\ell_1}\rangle_{11}\langle\varepsilon_{m+\ell_1}|
\otimes \sum_m
\varepsilon_{m,2}|\varepsilon_{m+\ell_2}\rangle_{22}\langle\varepsilon_{m+\ell_2}|
\otimes \cdots$ to be written as $\ve_{\ell_1,1}^{\prime} \otimes
\ve_{\ell_2,2}^{\prime} \otimes \cdots$ for all modes. Therefore,
when all $\ell_j$'s are zero, all detectors are positioned at the
position of $\hbar/m\omega$ on each coordinate. In the case where
$\ell_0=1$ while all other $\ell_j$'s are zero, the detector for
the first mode is positioned at $\lambda_{1}$ while all other
detectors are placed on $\lambda_0$'s. Similarly, when $\ell_0=2$,
the 2nd mode detector is placed at $\lambda_{2}$ and so on. When
all $\ell_j$'s are zero for detectors and $n_i$'s are zero for
particles, the measurement outcome for each detector would always
be zero, implying each particle is in the same position as the
corresponding detector, i.e., $\hbar/m\omega$. Suppose the first
particle is in the position of $\lambda_1$, then the first
detector would record the measurement outcome as $\Delta\lambda=
2\hbar/m\omega$ while the rest detectors would record zero.
However, suppose the second detector is positioned at $\lambda_2$,
i.e., the observable $\ve_{0,1}^{\prime} \otimes
\ve_{2,2}^{\prime} \otimes \ve_{0,3}^{\prime} \otimes \cdots$, and
the second particle positioned also at $\lambda_2$, i.e.,
$\ve_{0,1} \otimes \ve_{2,2} \otimes \ve_{0,3} \otimes \cdots $,
then all the detectors would still measure only zero.  With the
construction of reference frames in terms of position $\lambda$'s,
it provides a picture that when all particles are at $\lambda_0$
and all the detectors are also at $\lambda_0$, the measurement
outcomes will be zero. Therefore, while the energy of
electromagnetic field in vacuum diverges to infinity, the
detectors would measure only zero.

We have shown by using the example of a qubit measurement, that
the reference frame for the measurement can be constructed using
observables in such a way that it could be in one of the same
possible states as the qubit could be in. Moreover, the obtained
measurement outcomes are shown to be relative with respect to the
reference frame. We have applied this analogy to the case of a
single quantum harmonic oscillator and argued that just as with
the energy eigenvectors, the reference frame for a measurement can
only be in the non-zero ground level. This was then extended to
the electromagnetic field. It is noted that taking observables as
a reference frame for measurement has also yielded an interesting
result in regards to the halting problem in quantum computation
\cite{song}.

The author wishes to thank J. Bae for helpful discussions and J.M.
Isidro for correspondence on this topic.


\end{document}